\documentstyle[12pt,epsfig,amssymb]{article}
\newcommand{\arcosh}{{\rm arcosh}}

\def\lsim{\raise0.3ex\hbox{\,$<$\kern-0.75em\raise-1.1ex\hbox{$\sim$}\,}}
\def\gsim{\raise0.3ex\hbox{\,$>$\kern-0.75em\raise-1.1ex\hbox{$\sim$}\,}}

\begin{document}

\begin{titlepage}
\begin{flushright}
HIP-2004-29/TH\\ 
CERN-PH-TH/2004-112
\end{flushright}
\begin{centering}
\vfill

\vspace{-0.1cm}

{\bf THE FRAGILITY OF HIGH-$p_T$ HADRON SPECTRA \\ AS A HARD PROBE}

\vspace{0.5cm}
 K.J. Eskola${}^{\rm a,b,}$\footnote{kari.eskola@phys.jyu.fi},
 H. Honkanen${}^{\rm a,b,}$\footnote{heli.honkanen@phys.jyu.fi},
 C.A. Salgado${}^{\rm c,}$\footnote{carlos.salgado@cern.ch}
and  U.A. Wiedemann${}^{\rm c,}$\footnote{urs.wiedemann@cern.ch}\\

\vspace{1cm}
{\em ${}^{\rm a}$Department of Physics,\\
P.O.Box 35, FIN-40014 University of Jyv\"askyl\"a, Finland\\}
\vspace{0.3cm}
{\em ${}^{\rm b}$Helsinki Institute of Physics,\\
P.O.Box 64, FIN-00014 University of Helsinki, Finland\\}
\vspace{0.3cm}
{\em ${}^{\rm c}$Department of Physics, CERN, Theory Division,\\
CH-1211 Geneva 23, Switzerland\\}
\end{centering}

\vspace{1cm} \centerline{\bf Abstract} 
We study the suppression of high-$p_T$ hadron spectra in nuclear 
collisions, supplementing the perturbative QCD factorized formalism 
with radiative parton energy loss. We find that the nuclear modification
factor $R_{AA}(p_T)$, which quantifies the degree of suppression, is almost 
$p_T$-independent both for RHIC (in agreement with data) and for the LHC. 
This is a consequence of the shape of the partonic $p_T$-spectrum in 
elementary collisions which implies that for the same value of $R_{AA}$ 
at higher $p_T$, an increasingly smaller fraction of parton energy loss 
is needed. When the values of the time-averaged transport coefficient
$\hat q$ exceed $5\, {\rm GeV}^2/{\rm fm}$,  $R_{AA}(p_T)$ gradually 
loses its sensitivity to the corresponding produced energy density.
This is due to particle production in the outer corona of the medium, 
which remains almost unsuppressed even for extreme densities.
Thus, even for the highest experimentally 
accessible transverse momentum at the LHC and in contrast to jets, 
the measurement of leading 
partons via leading hadrons is not a penetrating probe of the dense 
matter, but a rather fragile probe which fragments for the 
opacities reached below the skin of the medium. 
Relating the transport coefficient
to the energy density produced in the collision region, we
find and discuss a phenomenon reminiscent of the opacity problem of 
elliptic flow: namely, the interaction of the hard parent parton with 
the medium appears to be much stronger than that expected for perturbative
interactions of the hard parton with an ideal quark gluon plasma.
\vfill
\end{titlepage}

\section{Introduction}

In Au+Au collisions at $\sqrt{s_{NN}} = 130$ and $200$ GeV, 
experiments at the Relativistic Heavy Ion Collider, RHIC, observe
a strong, a factor of 4-5, suppression of high-$p_T$ leading hadron spectra  
~\cite{Adcox:2001jp,Adler:2003au,Adler:2002xw,Adams:2003kv,Back:2003qr,
Arsene:2003yk}. The centrality dependence of this effect, the
corresponding suppression in leading back-to-back 
correlations ~\cite{Adler:2002tq}, and the moderate enhancement
and opposite centrality dependence observed in the same $p_T$-range in $d$+Au 
collisions~\cite{{Adler:2003ii},{Adams:2003im}} are strong indications
that the suppression is predominantly a final state effect
in which the energy of high-$p_T$ particles is degraded depending
on the in-medium pathlength and the density of the produced medium.
Formation time arguments~\cite{Wiedemann:2004wp} imply that for
$p_T > 5-7$ GeV, this energy loss is of partonic origin while 
hadronization occurs {\it outside} the partonic medium. 
This is also supported
by the disappearance of the particle species dependence of the 
suppression for $p_T > 5-7$ GeV.  
In this high-$p_T$ regime, the suppression of hadrons is widely 
regarded as a ``hard probe'' for the characterization of 
the energy density of the medium produced in nucleus-nucleus collisions. 
Recent studies, based on medium-induced radiative parton energy 
loss~\cite{Gyulassy:1993hr,Baier:1996sk,Zakharov:1997uu,Wiedemann:2000za,Gyulassy:2000er,Wang:2001if},
quote time-dependent energy densities which at time $\tau_0 = 0.2$ fm/c
after the collision lie a factor ${\cal O}(100)$ above the energy density of 
cold nuclear matter~\cite{Wang:2003aw,Gyulassy:2004vg}.

In this paper, we calculate the nuclear modification factor for 
high-$p_T$ hadronic spectra in nuclear collisions,
\begin{equation}
   R_{AA}(p_T,y)=\frac{d^2N^{AA}/dp_Tdy}{\langle T_{AA}\rangle_c
                 d^2\sigma^{NN}/dp_Tdy}\, .
   \label{eq1}
\end{equation}
Here, $\langle T_{AA}\rangle_c$ is the standard nuclear overlap 
function, calculated as the average in the measured centrality class.
In the absence of nuclear or medium effects, $R_{AA}\equiv 1$. The starting
point of our calculation is the collinearly factorized 
leading order (LO) perturbative QCD formalism which
reproduces the shape of the hadronic high-$p_T$ spectrum 
$d^2\sigma^{NN}/dp_Tdy$ in proton-proton collisions at RHIC
energies~\cite{Eskola:2002kv}. Our discussion will be limited
to $p_T\gtrsim$ 5 GeV. In the 
calculation of the charged hadron spectrum $d^2N^{AA}/dp_Tdy$
in this range, we include both initial and final state effects.
Nuclear modifications to parton distribution 
functions are taken from the EKS98 parametrization~\cite{Eskola:1998df} and 
isospin effects are included as in~\cite{Eskola:2002kv}. As a medium effect,
we consider partonic final state energy loss, which is introduced
via the quenching weights $P_f(\Delta E,L,\hat q)$ 
calculated in Ref.~\cite{Salgado:2003gb}. 
The quenching weights will be
discussed in more detail in section~\ref{sec2a}. They denote the
probability that a hard parton $f$ radiates an additional
medium-induced amount of energy $\Delta E$ if it propagates
through a medium of length $L$. The quenching weight depends on
the transport coefficient $\hat q$ which is determined by the 
average transverse momentum squared transferred to the projectile
parton per unit pathlength, $\hat q={\langle q^2_T\rangle}/{\lambda}$. 
It characterizes the energy density of the medium, 
$\hat{q} \propto \epsilon^{3/4}$~\cite{Baier:2002tc}.
With this input, the hadron spectrum in nucleus-nucleus collisions
reads schematically 
\begin{equation}
   d\sigma^{AA\to h+X}_{(\rm med)}
      =\sum_f d\sigma^{AA\to f+X}_{\rm (vac)}
       \otimes P_f(\Delta E,L,\hat q)
       \otimes D^{\rm (vac)}_{f\to h}(z,\mu^2_F)\, .
      \label{eq2}
\end{equation}
Here,  
\begin{equation}
    d\sigma^{AA\to f+X}_{\rm (vac)}
     =\sum_{ijk}f_{i/A}(x_1,Q^2)
     \otimes f_{j/A}(x_2,Q^2)\otimes\hat\sigma_{ij\to f+k}
     \label{eq3}
\end{equation}
and $f_{i/A}(x,Q^2)$ are the nuclear parton distribution functions 
and
$\sigma_{ij\to f+k}$ the perturbatively calculable partonic
cross sections.

The paper is organized as follows. In Section~\ref{sec2} we give details 
of the 
formalism used to compute hadron spectra with parton energy loss. 
Results for Au+Au collisions at $\sqrt{s_{\rm NN}}=200$ GeV are presented 
in Section~\ref{sec3} and predictions 
for the LHC are given in Section~\ref{sec4}. In Section~\ref{sec5}, we
discuss the nuclear modification factor for Au+Au collisions
at intermediate energy $\sqrt{s_{\rm NN}}=62.4$ GeV, where data 
are just being analyzed. In Section~\ref{sec6}, we discuss in detail
how the main model parameter of our analysis, the time-averaged
transport coefficient $\hat q$, is related to the energy density
produced in the medium. The main conclusions are summarized in
Section~\ref{sec7}. The results for RHIC and for the LHC are further explored
in Appendix~\ref{appa}.

\section{Framework}
\label{sec2}

\subsection{Quenching weights}
\label{sec2a}

Under the assumption that multiple gluons are emitted independent
of each other,
the probability that a parton loses an amount $\Delta E$
of its energy by medium-induced gluon radiation is~\cite{Baier:2001yt}, 
\begin{equation}
  P(\Delta E) = \sum_{n=0}^\infty \frac{1}{n!}
  \left[ \prod_{i=1}^n \int d\omega_i \frac{dI(\omega_i)}{d\omega}
    \right]
    \delta\left(\Delta E -\sum_{i=1}^n {\omega_i} \right)
    \exp\left[ - \int d\omega \frac{dI}{d\omega}\right]\ .
\label{eqqw}
\end{equation}
This probability distribution has been calculated and 
tabulated~\cite{Salgado:2003gb} for the medium-induced gluon energy 
distribution $\omega \frac{dI}{d\omega}$ of a hard parton.
[In the present study, we use a probability distribution calculated 
with $\alpha_S=0.5$ instead of $\alpha_S=1/3$. Since  
$\Delta E \propto \alpha_s\, \hat q $, the dependence on the value
of the coupling constant can be largely absorbed in a rescaling
of $\hat q$.]
The calculation of Ref.~\cite{Salgado:2003gb} was done in two approximations
which differ in viewing the medium as a source of many 
soft~\cite{Baier:1996sk,Zakharov:1997uu,Wiedemann:2000za} or a
few hard~\cite{Wiedemann:2000za,Gyulassy:2000er} momentum transfers,
respectively. For the purpose of the present study, the small
numerical differences between both approximations are not 
important~\cite{Salgado:2003gb}. We use the 
soft multiple scattering approximation, where the quenching weights 
$P(\Delta E)$ depend on the parton species (light quark or gluon),
the in-medium pathlength $L$ and the transport coefficient $\hat q$. 
For a static medium, $\hat q$ does not show a time-dependence. Extension and
density of the medium can be expressed in terms of the characteristic 
gluon frequency $\omega_c=\frac{1}{2}\hat qL^2$ and the dimensionless 
quantity $R=\omega_cL$.

\begin{figure}[h]
\begin{center}
\vspace{-1cm}
\epsfysize=9cm\epsffile{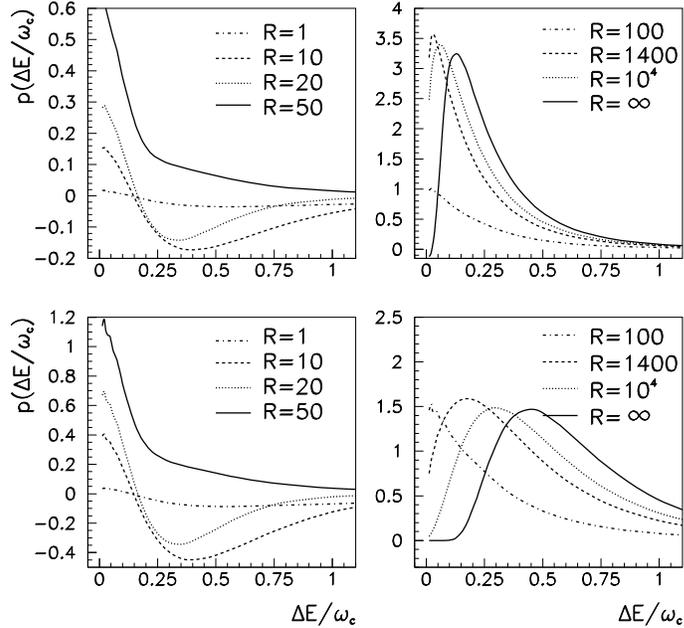}
\caption[a]{\protect \small Continuous part of the quenching weights 
for light quarks (upper row) and gluons (lower row)
computed in the multiple soft scattering 
approximation with $\alpha_S=0.5$.}
\label{fig1}
\end{center}
\end{figure}
%
In general, the density of the medium decreases strongly in time
due to longitudinal and transverse expansion. Hence, the transport
coefficient should be time-dependent. However, the gluon 
energy distribution $\omega \frac{dI}{d\omega}$ for a time-dependent 
transport coefficient $\hat{q}_{\tau}$ agrees with the energy 
distribution obtained for an {\it equivalent static} linear 
line-averaged transport coefficient~\cite{Salgado:2002cd,Baier:1998yf}
\begin{equation}
  \hat{q} = \frac{2}{L^2} \int_{\tau_0}^{\tau_0 + L}
  \tau\, \hat{q}_{\tau}(\tau)\, d\tau\, .
  \label{eq5}
\end{equation}
Here, $\tau_0$ is the time at which the parton is produced and 
for $\tau_0 \ll L$, the $\tau_0$-dependence of $\hat{q}$
is numerically negligible. This allows us to base our analysis
on an equivalent static scenario and to translate afterwards
with Eq.~(\ref{eq5}) the values for the time-averaged $\hat{q}$ 
into time-dependent transport coefficients .
%
\begin{figure}[h]
\begin{center}
\vspace{-1cm}
\epsfysize=7cm\epsffile{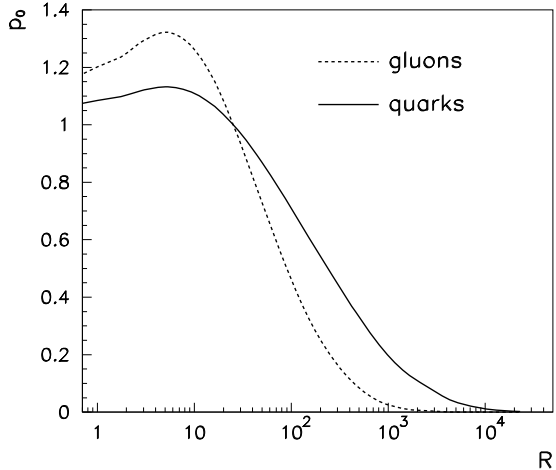}
\caption[a]{\protect \small Discrete part of the quenching weights
for light quarks  and gluons  computed in the multiple soft scattering
approximation with $\alpha_S=0.5$.}
\label{fig2}
\end{center}
\end{figure}

The probability distribution  $P(\Delta E/\omega_c,R)$ consists of a 
discrete and a continuous part (see Figs. \ref{fig1} and \ref{fig2})
\begin{equation}
  P(\Delta E/\omega_c,R)
  =p_0(R)\delta(\Delta E/\omega_c)+p(\Delta E/\omega_c,R)\, .
  \label{eq6}
\end{equation}
The continuous part denotes the probability to lose an additional energy
$\Delta E$ due to medium induced gluon radiation. The discrete contribution 
$p_0$ is the finite probability for the projectile to escape the collision 
region of finite size and finite opacity without further 
interaction~\cite{Salgado:2002cd}. The quenching weight 
$P(\Delta E/\omega_c,R)$ is normalized to 
unity,
\begin{equation}
  \int^\infty_0d(\Delta E/\omega_c)P(\Delta E/\omega_c,R)=p_0(R)+
  \int^\infty_0d(\Delta E/\omega_c)p(\Delta E/\omega_c,R)=1\, .
  \label{eq7}
\end{equation}
$P(\Delta E/\omega_c,R)$ is a generalized probability. It 
can take negative values since it describes a medium-induced 
redistribution of gluon radiation which - compared to the gluon 
radiation in the vacuum - can be {\it reduced} in some range of 
phase space~\cite{Wiedemann:2000tf}.

The quenching weights are computed in the eikonal limit of very large
initial parton energy $E_i$. For the case of realistic initial
parton energies, finite energy corrections have to be taken into
account. In the absence of a formalism which includes them a priori,
this is done by ensuring with kinematic cuts that the probability 
distribution $P(\Delta E/\omega_c,R)$ has weight only in the range
$\Delta E < E_i$~\cite{Baier:2001yt,Gyulassy:2001nm,Salgado:2003rv,Dainese:2003wq}. 
To illustrate the theoretical uncertainty associated with finite
energy corrections, we shall compare results without such corrections
to results obtained from the reweighted (`rw') 
probability~\cite{Salgado:2003rv}
\begin{equation}
   P_{\rm rw}(\Delta E/\omega_c,R)=\frac{P(\Delta E/\omega_c,R)}
   {\int^{E_i/\omega_c}_0d(\Delta E/\omega_c)P(\Delta E/\omega_c,R)}
   \Theta(1 - \Delta E/E_i)\, .
   \label{eqrw}
\end{equation}
We consider the non-reweighted result for the ratio $R_{AA}$
as a lower limit.

Numerically, we have checked that the effect of reweighting 
(\ref{eqrw}) is comparable to the finite energy corrections 
in other approaches~\cite{Gyulassy:2001nm} 
where kinematic cuts are imposed on the level of the gluon energy
distribution by multiplying with the vacuum splitting 
function.

\subsection{Hadron spectra with parton energy loss}
\label{sec2b}

We first discuss the nuclear geometry entering the calculation
of the nuclear modification factor (\ref{eq1}). 
The centrality dependence is taken into account by defining 
an effective nucleus $A_{\rm eff}$ for which the number of participants
in central $A_{\rm eff} + A_{\rm eff}$ collision equals 
$\langle N^{AA}_{\rm part}\rangle_c$. Here, the number of participants
is obtained from an optical Glauber analysis with realistic
Wood-Saxon (WS) profile for the centrality class $c$ \cite{Eskola:2001bf}.
As far as energy loss is concerned, the $A_{\rm eff}$ nucleus is 
assumed to be cylindrical and uniform, but we correct the final spectra
for the difference between $T^{\rm cyl}_{A_{\rm eff}A_{\rm eff}}(0)$ and 
$T^{\rm WS}_{A_{\rm eff}A_{\rm eff}}(0)$. 

The partons produced at midrapidity $y=0$ propagate in the transverse plane. 
Consider a parton which is produced at a position 
$\vec{s} = (s,\phi_S)$ with
respect to the transverse coordinates  and which propagates in
the transverse direction given by the angle
$\phi_L$. Its angle of propagation with respect to $\vec{s}$ is
$\phi_{LS}= \phi_L -\phi_S$. The geometrical 
transverse path length $L_{\rm geom}$ for this parton is 
\begin{equation}
  L_{\rm geom}(s,\phi_{LS}) =-s\cos\phi_{LS}+
    \sqrt{s^2{\cos\phi_{LS}}^2+R^2_{A_{\rm eff}}-s^2}\, ,
\end{equation}
where $R_{A_{\rm eff}}$ is the radius of $A_{\rm eff}$. We assume that
significant energy loss occurs only in the dense QGP-phase whose estimated 
lifetime $\tau_{\rm QGP}$ may be shorter than $L_{\rm geom}$. Therefore 
the in-medium path length which enters the quenching weight (\ref{eq5}) 
is chosen to be $L=\min(L_{\rm geom},L_{\rm cut}\sim\tau_{\rm QGP})$. 
Varying $L_{\rm cut}$ will turn out to be a useful tool to establish 
to what extent particle emission is surface dominated. 

We compute the inclusive production of charged particles at midrapidity 
in the framework
of factorized leading-order partonic cross sections, using
nuclear parton distributions, fragmentation functions and 
quenching weights but no intrinsic transverse momenta. As in 
\cite{Eskola:2002kv} the renormalization and factorization scales
are set to $\mu=Q=q_T$ and fragmentation 
scale to  $\mu_F=p_T$, where $q_T$ and $p_T$ are the transverse momentum of 
the parton and hadron, respectively. We use CTEQ5 parton 
distributions~\cite{Lai:1999wy} with nuclear effects for $A_{\rm eff}$
from EKS98~\cite{Eskola:1998df}, and KKP fragmentation 
functions~\cite{Kniehl:2000fe}. The $p+p$ reference spectra are obtained 
as in~\cite{Eskola:2002kv}. 

The inclusive cross section for the production of a  
parton of flavour $f$ and an initial transverse momentum $q_{T_i}$ and 
rapidity $y_i$ is given e.g. in Eqs. (4) and (5) of~\cite{Eskola:2002kv}. 
When traversing a medium, the parton will lose a fraction $\epsilon$ of 
its initial energy $E_i$, but it will not change significantly the  
direction of its trajectory, $\phi_i\equiv\phi_{LS}$. Therefore, we 
can write for the spectrum of the outgoing partons
\begin{eqnarray}
  \frac{dN^{AA\rightarrow f+X}}{d^2sdq_{T_f}d\phi_fdy_f} 
  &=&
   [T_{A}(s)]^2\int^1_0d\epsilon P(\epsilon,L,\hat q)\int d^2q_{T_i}dy_i
   \frac{d\sigma^{AA\rightarrow f+X}}{d^2q_{T_i}dy_i}
     \nonumber \\
  &&\times\delta(y_f-y_i)\delta(\phi_f-\phi_i)
    \delta(q_{T_f}-(1-\epsilon)q_{T_i})
    \nonumber \\
  &&\times\Theta(0\le\phi_i\le 2\pi)\nonumber \\
  &&\times\Theta\left(q_0\le q_{T_i}\le\frac{\sqrt{s}}{2\cosh{y_i}}
  \right)\Theta\left(\vert y_i\vert\le\arcosh\frac{\sqrt{s}}{2q_0}
  \right)\, ,
\label{fparton1}
\end{eqnarray}
where $T_{A}(s)=\frac{A}{(\pi R^2_{A})}\Theta (R^2_{A}-s^2)$, $q_0$ is 
the lower limit for the {\it initial} partonic transverse momentum and 
$P(\epsilon,L,\hat q)$ is the quenching weight. In our notation 
$\Delta E = \epsilon E_i =\epsilon\, q_{T_i} \cosh{y_i}$,
so 
\begin{equation}
  P(\epsilon,L,\hat q)=\frac{E_i}{\omega_c}P(\Delta E/\omega_c,R)
\end{equation}
and
\begin{equation}
  P_{\rm rw}(\epsilon,L,\hat q)=\frac{\frac{E_i}{\omega_c}
P(\Delta E/\omega_c,R)}
  {\int^{E_i/\omega_c}_0d(\Delta E/\omega_c)P(\Delta E/\omega_c,R)}.
\end{equation}
From Eq.~(\ref{fparton1}) we find that 
\begin{equation}
{\rm max}\left(0,1-q_{T_f}/q_0\right)\le\epsilon\le 
1-\frac{2q_{T_f}\cosh{y_f}}{\sqrt{s}}
\end{equation}
and 
\begin{eqnarray}
\frac{dN^{AA\rightarrow f+X}}
{dq_{T_f}^2dy_f} &=&
\int^{R_{A}}_0dss\int^{2\pi}_0d\phi_f\int d\epsilon
P(\epsilon,L,\hat q)[T_{A}(s)]^2
\frac{1}{(1-\epsilon)^2}
\frac{d\sigma^{AA\rightarrow f+X}}
{dq_{T_i}^2dy_i}\nonumber \\
&&\times\Theta\left(0\le q_{T_f}\le\frac{\sqrt{s}}{2\cosh{y_f}}
\right)\Theta\left(\vert y_f\vert\le\arcosh\frac{\sqrt{s}}{2q_0}
\right),
\label{fparton2}
\end{eqnarray}
with $q_{T_i}=q_{T_f}/(1-\epsilon)$ and $y_i=y_f$.

We define $z$ as the fraction of the energy of the parent parton carried 
by the leading hadron, $z=E_h/E_f$. For a hadron with transverse momentum 
$p_T$ we obtain~\cite{Eskola:2002kv}
\begin{eqnarray}
\frac{dN^{AA\rightarrow h+X}}{dp_T^2dy}
&=& K(\sqrt s)\cdot J(m_T,y)
\sum_f\int\frac{dz}{z^2}\, 
D_{f\rightarrow h}(z,\mu_F^2) 
\frac{dN^{AA\rightarrow f+X}}{dq_{T_f}^2 dy_f}\, ,
\label{eqcas13}
\end{eqnarray}
where
\begin{equation}
q_{T_f}=\frac{p_T}{z}J(m_T,y),\quad y_f={\rm arsinh}\left(\frac {m_T}{p_T}\sinh y\right)
\end{equation}
and 
\begin{equation}
J(m_T,y) = \left(1-\frac{m^2}{m_T^2\cosh^2 y}\right)^{-1/2}.
\label{jaakoppi}
\end{equation}
The integration region for $z$ is 
\begin{equation}
\frac{2m_T}{\sqrt s}\cosh{y}\le z\le 1\, ,
\end{equation}
since 
we do not attempt to explore the region $p_T <q_0$. The 
cms-energy dependent $K$-factors applicable within our framework
were determined in~\cite{Eskola:2002kv}. 
They are important for the overall normalization of the spectrum but
they cancel in the nuclear modification factor $R_{AA}$ discussed
in the next sections.

Our formalism is equivalent to defining medium-modified fragmentation 
function \cite{{Salgado:2002cd},{Gyulassy:2001nm},{Wang:1996yh}}
\begin{equation}
  D^{({\rm med})}_{f\rightarrow h}(z,\mu_F^2)=\int d\epsilon P(\epsilon)
  \frac{1}{(1-\epsilon)}
  D_{f\rightarrow h}\left(\frac{z}{1-\epsilon},\mu_F^2\right). 
\end{equation} 
Neglecting the masses, Eq.~(\ref{eqcas13}) simplifies~\cite{Baier:2001yt}
for the case of small $\Delta E$ and $z=E_h/E_f\sim 1$ to
\begin{equation}
\frac{dN^{AA\rightarrow h+X}}{dp_Tdy}
\sim \int d(\Delta E)P(\Delta E) \frac{dN^{\rm (vac)}(p_T+\Delta E)}{dp_T dy}\, .
\label{eqcasqf}
\end{equation}
Here, $dN^{\rm (vac)}/dp_T dy$ is the spectrum of hadrons in 
the case
of no medium (i.e. it coincides with the proton-proton spectrum in the absence
of initial state effects as shadowing etc.).
The suppression computed with (\ref{eqcasqf}) gives a rather good
approximation to the one computed with the full formula (\ref{eqcas13}).

\section{Results for RHIC at $\sqrt{s_{\rm NN}} = 200$ GeV}
\label{sec3}
Based on the above formalism, we have studied the nuclear modification
factor $R_{AA}(p_T)$ for charged hadrons, $h\equiv (h^++ h^-)/2$ in
Au+Au collisions at $\sqrt{s_{NN}} = 200$ GeV. We include charged
pions, kaons and (anti)protons in the sum of charged particles. Our 
discussion is limited to the $0-5\%$ most central collisions. This
centrality class corresponds to the head-on collision of an effective 
nucleus of size $A_{\rm eff}=181$. 

In Fig.~\ref{fig3}, we compare data from RHIC with the 
nuclear modification factor calculated for different values of the 
transport coefficient $\hat q$. This figure is for a medium which
exists for at most $\tau\sim L_{\rm cut} = 5$ fm in the dense partonic phase.
We find that even in the absence of final state parton energy loss,
$\hat q = 0$, the nuclear modification factor
shows a reduction from the baseline $R_{AA}\equiv 1$ for $p_T \gsim 13$ GeV.
The reason is that in this $p_T$ region, the incoming
parton momentum fractions 
$x\simeq\frac{2p_T/\langle z\rangle}{\sqrt{s}}$ sampled by
the hadronic spectrum lie in the so-called EMC region where 
nuclear parton distribution functions decrease faster than free proton
parton distribution functions with increasing 
$x$~\cite{Eskola:1998df}. This decrease of the incoming parton
flux translates into a $p_T$-dependent decrease of the nuclear
modification factor $R_{AA}$. As the density of the medium
(or equivalently the transport coefficient) is increased, the
nuclear modification factor is reduced further. This is
caused by parton energy loss in the final state. 
Interestingly, the resulting nuclear modification factor $R_{AA}$
is almost $p_T$-independent and its dependence on $\hat q$ becomes
weaker as $\hat q$ increases. We now discuss these two features
in more detail:

\begin{figure}[h]
\centerline{\hspace{-2.cm} 
\epsfysize=9cm\epsffile{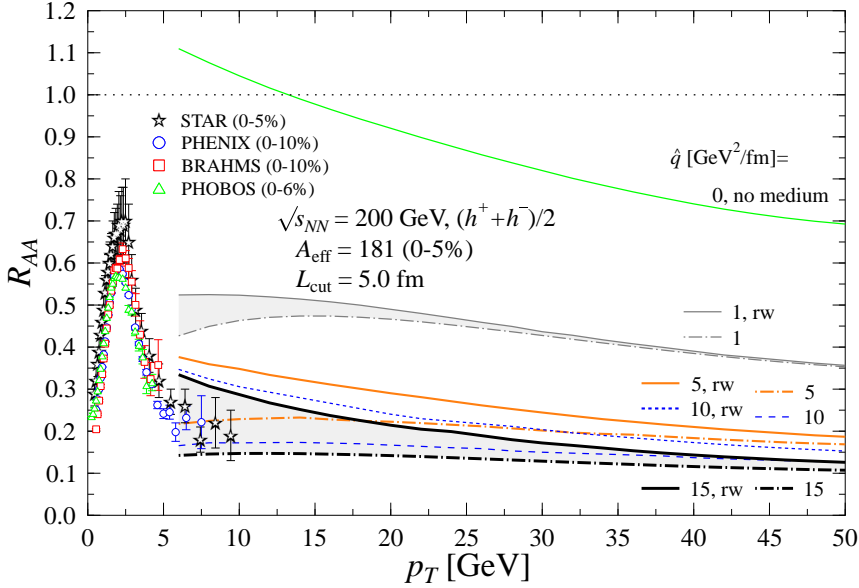}} 
\caption[a]{\protect \small (Colour online)
The nuclear modification factor 
$R_{AA}$ for charged hadrons $h\equiv (h^++ h^-)/2$
in the $0-5\%$ most central Au+Au collisions 
at $\sqrt{s_{NN}} = 200$ GeV. Different lines are for 
a maximal lifetime of the medium $\tau\sim L_{\rm cut}$=5 fm and
different values of the transport coefficient $\hat{q}$.
The shaded region between the curves calculated
with reweighted (solid curves) and non-reweighted (dotted-dashed curves) 
quenching weights is indicative of uncertainties related to finite
energy corrections. For comparison, 
the most central $R_{AA}$ data from the RHIC collaborations are 
shown (with statistical errors only): 
$0-5\%$ STAR data for $\vert\eta\vert <0.5$ \cite{Adams:2003kv},
$0-10\%$ PHENIX data 
for $\vert\eta\vert<0.35$ \cite{Adler:2003au},
$0-10\%$ BRAHMS data for $\eta=0$ \cite{Arsene:2003yk}
and $0-6\%$ PHOBOS data for $0.2<\eta<1.4$ \cite{Back:2003qr}.}
\label{fig3}
\end{figure}

To understand the approximate $p_T$-independence of $R_{AA}$, we study
the sensitivity of $R_{AA}$ to the $p_T$-dependent 
shape of the partonic 
cross section entering (\ref{eq2}). (For qualitative
results we assume $z\sim 1$, which is a fairly good approximation
for quark dominated hadron production at RHIC
in the large-$p_T$ region of our interest.)
We consider first a partonic cross 
section which has a powerlaw tail $d\sigma/dp_T\sim 1/p_T^n$,
$n$ constant. If $n$ is sufficiently
large, then the medium-modified spectrum can be obtained by shifting the
vacuum spectrum by $S(p_T) \sim \sqrt{p_T/n}$~\cite{Baier:2001yt}, and
one finds $R_{AA}(p_T) \sim \left(1 + c/\sqrt{n p_T}\right)^{-n}$
where $c = \sqrt{2\pi \alpha_s^2 \omega_c}$. In this case, $R_{AA}$
approaches unity in the limit of large $p_T$, 
except for possible modifications in the incoming parton distributions.
In contrast, if the partonic spectrum is exponential, one finds a 
$p_T$-independent nuclear modification factor
\begin{equation}
   R_{AA}(p_T)
       \sim \int d(\Delta E)P(\Delta E) \frac{\exp\{-a(p_T+\Delta E)\}}
       {\exp\{-a p_T\}}=\int d(\Delta E)P(\Delta E)\exp\{-a\Delta E\}\, .
   \label{eqcasqff}
\end{equation}
For the spectra of produced quarks and gluons computed from 
Eq.~(\ref{eqcas13}) at RHIC energies, one finds that the 
logarithmic slope $n$ of the partonic $p_T$-spectrum 
increases with increasing $p_T$. This makes parton energy
loss more effective in reducing $R_{AA}$ at high $p_T$. Moreover, 
due to kinematic boundary effects, the underlying partonic spectrum 
at RHIC can be parametrized by an exponential 
$\exp\{-a p_T\}$ for $p_T\gtrsim$ 30 GeV. As a consequence of
(\ref{eqcasqff}), one thus finds - in agreement with
Fig.~\ref{fig3} - an almost $p_T$-independent $R_{AA}$
at large $p_T \gtrsim 30$ GeV. [Although this is not within 
experimental reach, we note that close to the kinematical limit 
$ p_T \sim 100$ GeV, the slope decreases even faster.] 
The remaining weak $p_T$-dependence
can be attributed to the parton species dependence 
of the quenching weights and to
the nuclear dependences of the incoming parton 
distribution function which turn out to further reduce $R_{AA}$ 
with increasing $p_T$.
In conclusion, the onset of a perturbative $p_T$ powerlaw 
(at $p_T > 5$ GeV say), and its return to an exponential spectrum at
$p_T \gtrsim 30$ GeV leave a finite window at RHIC energies. Within
this window, the gradually increasing slope of the partonic spectrum 
implies an almost $p_T$-independent nuclear modification factor. 

\begin{figure}[h]
\vspace{-1.cm}
\centerline{\epsfysize=9cm\epsffile{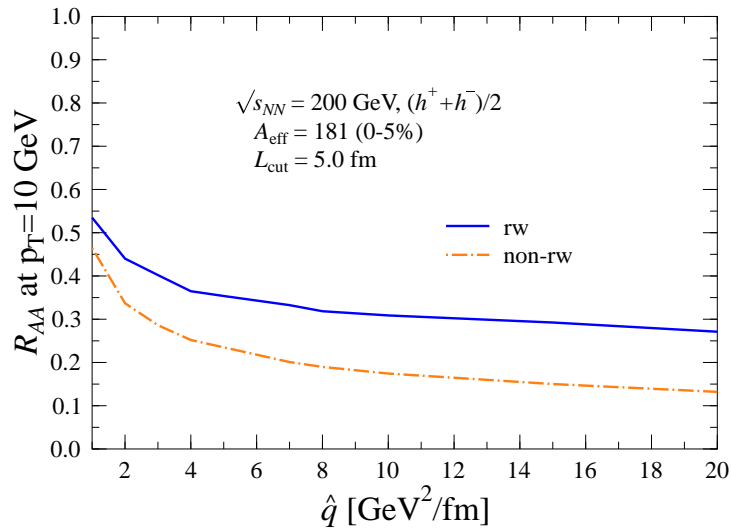}}
\caption[a]{\protect \small (Colour online)
$R_{AA}$ for charged hadrons $h\equiv (h^++ h^-)/2$
in the $0-5\%$ most central Au+Au collisions 
at $\sqrt{s_{NN}} = 200$ GeV
as a function of $\hat q$ for $p_T$=10
GeV.}
\label{fig4}
\end{figure}

We now turn to the dependence of $R_{AA}$ on $\hat q$.
Partons which traverse a larger in-medium pathlength tend to lose
more energy. As a consequence, the dominant fraction of leading
partons comes from the outer skin of the two-dimensional transverse
overlap of the colliding nuclei. We refer to the outer region of
the medium as ``skin'' rather than ``surface'' since it has a 
finite three-volume. In Appendix~\ref{appa}, we give further support
for this dominance of skin-emission for realistic densities at RHIC
and at the LHC.
The dominance of skin-emission appears to limit the
sensitivity of $R_{AA}$ on the density of the medium. Indeed,
we find that for sufficiently large but realistic densities of 
the medium, the thickness of the outer skin changes only weakly 
with increasing density, see Fig.~\ref{fig4}. Thus
the nuclear modification 
factor is a sensitive measure of the density of the medium up to
$\hat q \lsim 4\ {\rm GeV}^2/{\rm fm}$, but it loses this sensitivity 
for higher values of $\hat q$.

From Fig.~\ref{fig3} and~\ref{fig4} we conclude
that the suppression of high-$p_T$ hadrons measured in 
$\sqrt{s_{\rm NN}} = 200$ GeV Au+Au collisions at RHIC 
requires a time-averaged transport coefficient $\hat q > 5$ 
${\rm GeV}^2/{\rm fm}$ and favours
$\hat q=5-15$ ${\rm GeV}^2/{\rm fm}$.

\section{Extrapolation to the LHC}
\label{sec4}
Within the formalism of section~\ref{sec2}, the choice of the
transport coefficient $\hat q$ is the main unknown for
extrapolating our results from RHIC to Pb+Pb collisions at the 
LHC energy $\sqrt{s_{NN}}=5500$ GeV. One expects that this
transport coefficient grows linearly with the number density, $n_i$, of the
medium~\cite{Baier:2002tc,Baier:1998yf}, and hence is proportional 
to the multiplicity per unit rapidity
\begin{equation}
  \hat{q} \propto \frac{dN^{\rm ch}}{dy}\, .
  \label{eq22}
\end{equation}
For the LHC, model predictions of the event multiplicity vary by 
a factor $\sim 4$ at least~\cite{Armesto:2000xh}, but the corresponding 
uncertainty for $R_{AA}$ may be significantly smaller since $R_{AA}$ 
changes only weakly for large values of $\hat{q}$. The second unknown 
in the calculation of $R_{AA}$ 
is the maximal lifetime of the dense medium, denoted by $L_{\rm cut}$.
We expect, however, that this choice is numerically unimportant 
since $L_{\rm cut}$ at LHC is typically larger than the thickness of the
effective
skin of the medium through which hard partons can penetrate without
significant energy loss. 

\begin{figure}[h]
\centerline{\hspace{-2.cm} 
\epsfysize=9cm\epsffile{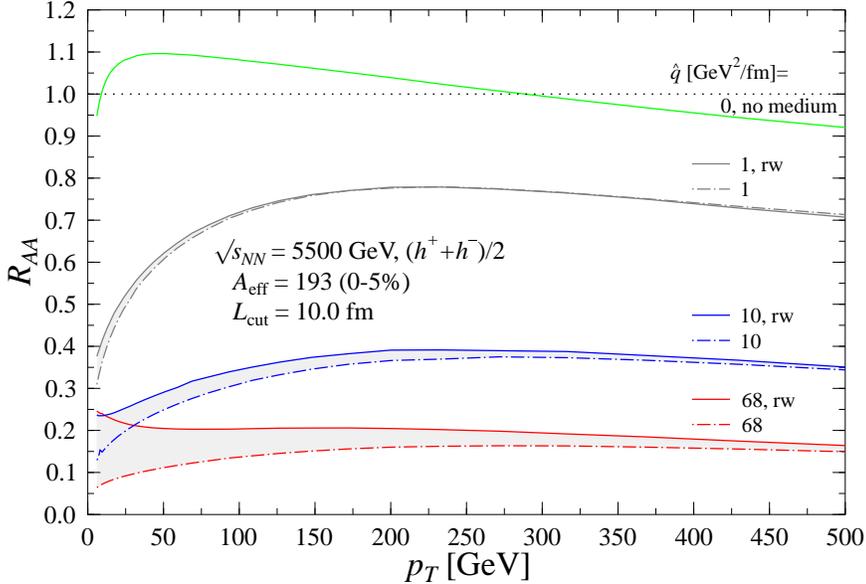}} 
\caption[a]{\protect \small (Colour online)
The nuclear modification factor 
for charged hadrons $h\equiv (h^++ h^-)/2$ in 
$0-5\%$ most central Pb+Pb
collisions at the LHC, calculated for $L_{\rm cut}$=10 fm and 
various values of the transport coefficient. The largest value
$\hat q_{\rm LHC}=68$ ${\rm GeV}^2/{\rm fm}$ lies in the range
favoured by current multiplicity estimates. See the text for more details.}
\label{fig5}
\end{figure}

To be specific, we use for the LHC $L_{\rm cut} = \tau_{\rm QGP}\sim$ 
10 fm consistent with hydrodynamical modeling of the time evolution 
of the medium~\cite{EHNRRT}. We find from an optical Glauber calculation that
the $0-5\%$ most central Pb+Pb events correspond to the head-on
collision of nuclei with $A_{\rm eff}=193$. The density estimate
is obtained from the EKRT parametrization \cite{Eskola:1999fc} 
for which $n_i\sim A^{0.383}(\sqrt{s})^{0.574}$, consistent with
the growth of the event multiplicity measured so far. With this
input, the transport coefficients at RHIC and the LHC are related
like $\hat q_{\rm LHC}=6.8 \hat q_{\rm RHIC}$. For a typical RHIC
value, $\hat q_{\rm RHIC} = 10 $ ${\rm GeV}^2/{\rm fm}$, we find
that $\hat q_{\rm LHC}=68$ ${\rm GeV}^2/{\rm fm}$ which is one of
the parameters used in Fig.~\ref{fig5}. In models which predict a
smaller multiplicity increase from RHIC to LHC~\cite{Armesto:2000xh}, 
the suppression is expected to lie in between the bands shown in
Fig.~\ref{fig5} for $\hat q_{\rm RHIC} = 10 $ ${\rm GeV}^2/{\rm fm}$
and $\hat q_{\rm LHC}=68$ ${\rm GeV}^2/{\rm fm}$.

Based on estimates of the luminosity in Pb+Pb collisions at the
LHC and the hadronic cross section, one expects that the nuclear
modification factor will be measured up to $p_T \sim 100 - 150$
GeV. Not only in this $p_T$-range but even above, the nuclear
modification factor turns out to be almost $p_T$-independent
for $\hat q_{\rm LHC}=68$ ${\rm GeV}^2/{\rm fm}$. Moreover, 
it is numerically comparable to $R_{AA}$ at RHIC, calculated
for a much smaller transport coefficient. 
This can be understood as follows: the contribution from the discrete 
part of the quenching weight to $R_{AA}$ decreases with the 
increasing transport coefficient from 
RHIC to the LHC (see Fig.~\ref{fig2} and Appendix~\ref{appa}). 
At the same time, however, since the
partonic transverse momentum spectra are steeper at RHIC than at the 
LHC, the suppression of $R_{AA}$ induced by the continuous part of the 
quenching weight (parton energy loss) is still larger at RHIC than at 
the LHC (see discussion in Appendix~\ref{appa}).
If the density of the medium created at the LHC 
is significantly reduced ($\hat q_{\rm LHC}=10$ ${\rm GeV}^2/{\rm fm}$),
then one finds a weak but experimentally accessible increase of 
$R_{AA}(p_T)$ with $p_T$. The reason is that for sufficiently
low density, the probability that partons from deeper layers
inside the medium escape with small energy loss grows significantly
with increasing $p_T$. This effect is even more pronounced for
very small densities of order $\hat q_{\rm LHC}=1$ ${\rm GeV}^2/{\rm fm}$.

From this exercise, we expect that up to the highest transverse
momenta accessible at the LHC, leading hadron spectra will show
a strong suppression in nuclear collisions. This suppression is
likely to result in an almost $p_T$-independent nuclear modification
factor. For the reasons given above and since particle emission from 
the outer skin of the medium remains essentially unsuppressed even 
for extreme densities, we expect that the numerical value for 
$R_{AA}$ at the LHC is comparable to the one at RHIC.

\section{The nuclear modification factor at  
$\sqrt{s_{\rm NN}} = 62.4$ GeV}
\label{sec5}

\begin{figure}[h]
\vspace{-1.cm}
\centerline{
\epsfysize=9cm\epsffile{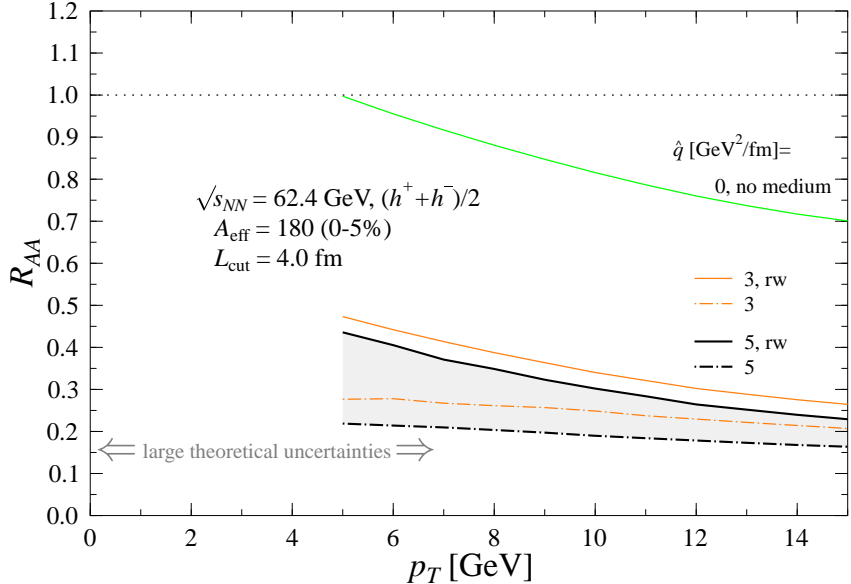}}
\caption[a]{\protect \small (Colour online)
The nuclear modification factor
for charged hadrons $h\equiv (h^++ h^-)/2$ in $0-5\%$ most
central Au+Au collisions at $\sqrt{s} = 62.4$ GeV. See the text for details.}
\label{fig6}
\end{figure}

Experiments at the Relativistic Heavy Ion Collider RHIC have just
completed data taking for Au+Au collisions at an intermediate center 
of mass energy of $\sqrt{s_{\rm NN}} = 62.4$ GeV. In 
Fig.~\ref{fig6}, we give results for the corresponding
nuclear modification factor in the present formalism. The main model 
input is again the value of the time-averaged transport coefficient 
$\hat{q}$ defined in (\ref{eq5}).
Based on the multiplicity scaling of the transport coefficient in
(\ref{eq22}) and the value $\hat q \sim 5 - 10\, {\rm GeV}^2/{\rm fm}$
favoured for RHIC data at $\sqrt{s_{\rm NN}} = 200$ GeV, we chose
$\hat q \sim 3 - 5\,  {\rm GeV}^2/{\rm fm}$
at $\sqrt{s_{\rm NN}} = 62.4$ GeV. For the second model-dependent
input, the life-time $L_{\rm cut}$ of the medium,
we chose a slightly reduced value $L_{\rm cut} = 4$ fm. However the
dependence of our results on $L_{\rm cut}$ turns out to be negligible
for sufficiently large $L_{\rm cut}$. For medium densities favoured by 
multiplicity scaling the suppression is pronounced.
The results shown in Fig.~\ref{fig6} are,
within theoretical uncertainties, consistent with the results
published in several recent 
studies~\cite{Vitev:2004gn,Adil:2004cn,Wang:2004yv}. 

%


Some caveats should be kept in mind.
First, at intermediate $p_T < 5-7$ GeV, other physical
mechanisms can provide sizeable contributions to $R_{AA}$. In
particular, the anomalous baryon enhancement at intermediate
$p_T$ ~\cite{Adler:2003kg,Adams:2003am} and the Cronin effect~\cite{dA} 
are observed to increase $R_{AA}$.
Both contributions may be more pronounced on a spectrum with
steeper slope, i.e. for smaller $\sqrt{s_{\rm NN}}$. Second,
the smaller the hadronic $p_T$, the larger the theoretical 
uncertainty in calculating parton energy loss. Moreover, below
$p_T \sim 4$ GeV, the hadron spectra at $\sqrt{s_{\rm NN}} = 62.4$ GeV
deviate significantly from the shape calculated by the collinear
factorized pQCD formalism~\cite{Eskola:2002kv}.

While completing this work, the first experimental data on the 
nuclear modification factor in Au+Au collisions at 
$\sqrt{s_{\rm NN}} = 62.4$ GeV appeared~\cite{Back:2004ra}. These 
data are for $0.2<\eta<1.4$ and $p_T \le 4$ GeV with
$R_{AA}(p_T \sim 4 {\rm GeV}) \sim 0.8 \pm 0.1$ for the $0-6\%$ 
most central bin. Other preliminary data~\cite{pAprelim} indicate
a significant further drop of $R_{AA}$  for $4 < p_T < 6.5$ GeV.
For the highest $p_T$, these data are close to the calculation
shown in Fig.~\ref{fig6}.

\section{Relating the Transport Coefficient to Energy Density}
\label{sec6}

In section~\ref{sec3}, we have shown that RHIC data favour
a time-averaged transport coefficient 
$\hat q \simeq 10\ {\rm GeV}^2/{\rm fm}$. 
This parameter as well as the time-dependent transport coefficient entering
eq. (\ref{eq5}) scales with the energy density of the medium 
like~\cite{Baier:2002tc} 
\begin{equation}
 \hat q_\tau(\tau)= c\,  \epsilon^{3/4}(\tau)\, .
 \label{eq23}
\end{equation}
Here, $c$ is a proportionality constant which can be calculated
for specific models of the medium. In particular, for the model 
of an ideal quark gluon
plasma whose constituents interact perturbatively with the hard
parton, we extract this proportionality constant from Figure 3 
of Ref.~\cite{Baier:2002tc}  
%
\begin{equation}
  c_{\rm QGP}^{\rm ideal} \approx 2\, .
  \label{eq24}
\end{equation}
Alternatively, one can determine the constant $c$ from experiment
if one has independent information about transport coefficient and
energy density. From the measured time-averaged transport coefficient
$\hat q$, we find via the dynamical scaling law (\ref{eq5}) the
transport coefficient for an initial time $\tau_0$,
\begin{equation}
  \hat{q}_\tau (\tau_0) = \hat{q} \frac{2-\alpha}{2} 
  \left( \frac{L}{\tau_0} \right)^{\alpha}\, .
  \label{eq25}
\end{equation}
This expression is for an expanding medium with $\hat q_\tau(\tau)
= \hat q_\tau(\tau_0)\, \left(\tau_0/L \right)^\alpha$. We have assumed 
$\tau_0 \ll L$. The expansion parameter $\alpha$ is unity for a
one-dimensional Bjorken expansion and is expected to stay close 
to unity for realistic expansion scenarios. 
From this we find
\begin{eqnarray}
  c &=& \frac{\hat q_\tau (\tau_0)}{\epsilon(\tau_0)^{3/4}}
    = \frac{\hat q}{\epsilon(\tau_0)^{3/4}}
      \frac{2-\alpha}{2} 
  \left( \frac{L}{\tau_0} \right)^{\alpha}
\end{eqnarray}
For a typical initial time $\tau_0 \sim 0.2$ fm/c, the average in-medium
pathlength can be expected to be $\sim 10\, \tau_0 = 2$ fm 
(see Appendix~\ref{appa}).
The energy density averaged over a uniform transverse profile
at initial time $\tau_0$ can be as large as $\epsilon(\tau_0)\sim{\cal O}(100 
\ \frac{\rm GeV}{{\rm fm}^3})$, as suggested by 
the pQCD+saturation+hydrodynamics 
model \cite{Eskola:1999fc} which correctly predicted the final 
multiplicities and transverse energies in central collisions at
RHIC \cite{Eskola:2001bf}.
This implies that realistic scenarios lie in the
parameter range $\epsilon(\tau_0) < 100\, \frac{\rm GeV}{{\rm fm}^3}$,
$L\sim 10\, \tau_0$, $0.75 < \alpha < 1.5$, for which we find
\begin{eqnarray}
   c> 8\ldots 19\, .
   \label{eq26}
\end{eqnarray} 
Thus, for the entire realistic parameter
range, we find a proportionality constant which is 
a factor $\sim 4 - 5$
larger than the perturbative estimate. We note
that $\hat q$ is the product of the density of scattering centers
times their elastic scattering cross section~\cite{Baier:2002tc}.
Thus, the factor  $\sim 4 - 5$ discrepancy observed here is
reminiscent of the opacity problem in elliptic flow, where also
scattering cross sections much larger than the perturbative ones
are favoured by the data~\cite{Molnar:2001ux}. 

\section{Conclusions}
\label{sec7}

In this paper we have studied within the LO pQCD 
formalism~\cite{Eskola:2002kv} the effect of medium-induced
parton energy loss on charged hadron spectra. At high $p_T$ we find
generically an almost $p_T$-independent nuclear modification
factor at both LHC and RHIC energies. This is a direct 
consequence of the interplay of medium-induced energy
loss and of a partonic $p_T$-spectrum whose slope {\it increases} 
with increasing $p_T$. In this approach, an almost $p_T$-independent
$R_{AA}$ is obtained without fine-tuning of 
nuclear effects (such as Cronin or anomalous baryon
enhancement).

For central Au+Au collisions at $\sqrt{s_{NN}} = 200$ GeV, 
we find a satisfactory agreement with experiment for a transport
coefficient in the range $\hat q\sim 5 - 15$ GeV$^2$/fm. 
For these large values of $\hat q$, the nuclear modification factor 
is only weakly dependent on the transport coefficient, which makes
it difficult to constrain the energy density with
$R_{\rm AuAu}$, see Fig.~\ref{fig4}. On the other hand, this 
weakened sensitivity of $R_{AA}$ makes our prediction for the nuclear 
suppression at the LHC more stable. Irrespective of the uncertainty in the
factor by which the density of the medium increases from RHIC to the LHC, 
the nuclear modification factor is expected to lie in a narrow band 
$R_{\rm PbPb} \sim 0.2 \pm 0.1$. This and the almost $p_T$-independent
slope of $R_{\rm PbPb}$ are the main predictions of the present work
for Pb+Pb collisions at the LHC. Thus, our calculations indicate
that even in a kinematic regime $p_T \sim 100$ GeV where the same
energy loss formalism results in jet cross sections which are almost 
unaffected by the medium~\cite{Salgado:2003rv}, the suppression of 
leading hadron spectra is still sizeable. This points to the importance 
of resolving the internal
structure of jets, 
i.e. the jet shape and jet multiplicity distributions,
at the LHC in order to access the medium-dependence.

In section~\ref{sec6}, we have inverted the standard logic of 
the analysis of nuclear modification factors. Rather than to
determine the energy density $\epsilon$ from the measured transport 
coefficient, we extract the proportionality factor 
$c = {\hat q}\,\epsilon^{-3/4}$ for realistic 
upper limit estimates of the
energy density. This proportionality factor characterizes the
strength of the interaction between the hard (test) parton and the 
medium. It turns out to be at least a factor $\approx 4 - 5$
larger than 
expected from perturbative estimates [see discussion of 
eq. (\ref{eq24})]. Although this perturbative estimate (\ref{eq24})
is itself subject to significant uncertainties, we do not expect
that this is the
only source of such a large discrepancy. Here, we note that our
finding is reminiscent of parton cascade studies of elliptic 
flow~\cite{Molnar:2001ux} where perturbative partonic 
cross sections have to be scaled by large factors in order to match 
the size of the collective evolution. Both effects may have a common 
origin since both concern the interaction strength of partonic
degrees of freedom with the medium produced in the collision.  

{\it Note added: }
While completing this work, a calculation of nuclear modification
factors in a closely related model appeared ~\cite{Dainese:2004te}. 
This study supports the approximate $p_T$-independence of the
nuclear modification factor at RHIC and at the LHC. Moreover, this
model calculation reproduces the observed centrality dependence of
$R_{AA}$ and of the suppression of the away-side jet-like 
correlations. That the centrality dependence of suppressed high-$p_T$
hadroproduction follows naturally, once the suppression for the
most central bin is fixed, was also found in Ref.~\cite{Drees:2003zh}.

\subsection*{Acknowledgements}
We thank N. Armesto, R. Baier, A. Dainese and P.V. Ruuskanen
for helpful discussions. 
Financial support from the Academy of Finland, the projects 50338 and 
206024, is gratefully acknowledged.

%
\appendix

\section{Further studies of the dependences of $R_{AA}$}
\label{appa}
\begin{figure}[h]
\centerline{\hspace{-2.cm} 
\epsfysize=9cm\epsffile{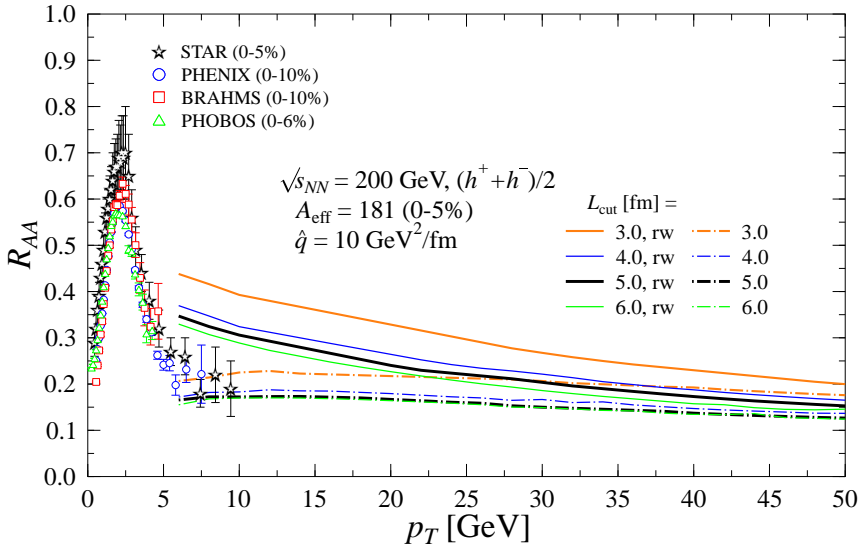}} 
\caption[a]{\protect \small (Colour online)
The nuclear modification factor 
$R_{AA}$ for charged hadrons $h\equiv (h^++ h^-)/2$
in the $0-5\%$ most central Au+Au collisions 
at $\sqrt{s_{NN}} = 200$ GeV. The solid curves (dotted-dashed curves)
are the reweighted (non-reweighted) results
with $\hat q=10$ ${\rm GeV}^2/{\rm fm}$ and
$L_{\rm cut}$=3, 4, 5 and 6 fm. The data are the same as in Fig.~\ref{fig3}.}
\label{fig7}
\end{figure}

We have argued in section~\ref{sec3} that for the densities attained
at RHIC, leading hadroproduction is limited to the outer skin of
the collision region. Here, we further support
this argument by calculating the dependence of $R_{AA}$ on $L_{\rm cut}$.
In particular, Fig.~\ref{fig7} establishes that a reduction of 
the maximum in-medium pathlength from 6 fm to 4 fm leaves the nuclear 
modification factor almost unaffected. The slight sensitivity of $R_{AA}$ 
to $L_{\rm cut}$=3 fm indicates that a small  
contribution comes from partons which traverse $3-4$ fm of matter
with transport coefficient $\hat q=10$ ${\rm GeV}^2/{\rm fm}$.
Thus, essentially all partons with $L > 3$ fm, as well as a fraction
of the partons with $L < 3$ fm do not contribute to the leading hadron
spectrum. The average geometrical path length is larger than $L = 3$ fm,
it is $\langle L_{\rm geom}\rangle =\frac{8}{3\pi}R_A = 5.2$ fm
for a cylindrical uniform nucleus 
(we use $A_{\rm eff}=181$ and $R_{A_{\rm eff}}=6.2$). 
Thus, much less than half of the partons will succeed in escaping the 
medium with sufficient energy, in agreement with
the observed value for $R_{\rm AuAu}$ at RHIC. 

\begin{figure}[h]
\vspace{-0.5cm}
\centerline{ 
\epsfysize=7cm\epsffile{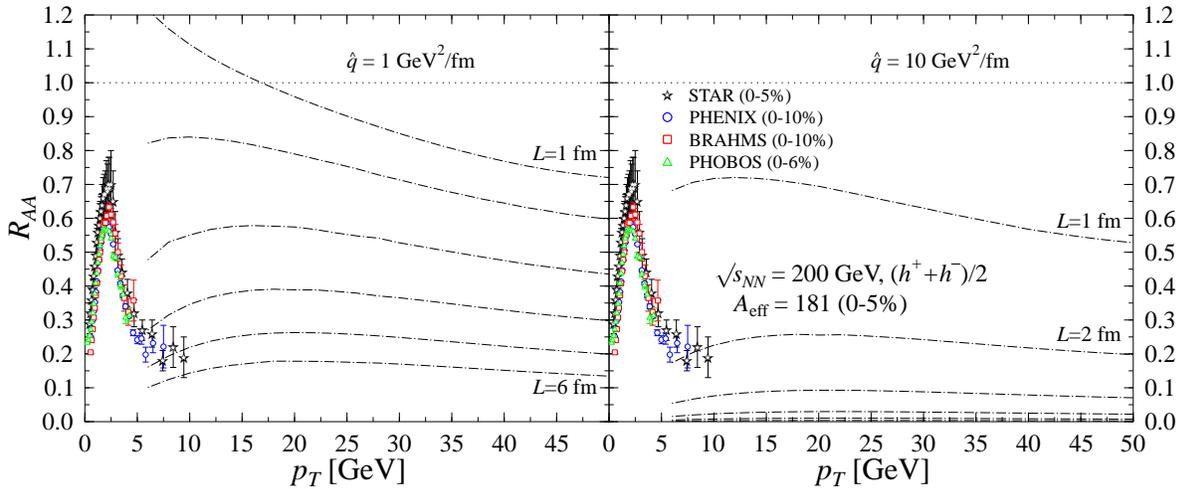}}
\caption[a]{\protect \small (Colour online)
The non-reweighted 
$R_{AA}$ results at RHIC computed using constant 
in-medium path length $L$ for $\hat q=1$  ${\rm GeV}^2/{\rm fm}$ (left panel)
and $\hat q=10$  ${\rm GeV}^2/{\rm fm}$ (right panel). See the text
for further details. The data are the same as in Fig.~\ref{fig3}.}
\label{fig8}
\end{figure}

Another illustration of skin-dominated leading hadroproduction in
nucleus-nucleus collision is to compare $R_{AA}$ calculated for
several {\it fixed}
in-medium pathlengths $L=1...6$ fm, see Fig.~\ref{fig8}.
For a small transport coefficient, $\hat q=1$ ${\rm GeV}^2/{\rm fm}$,
$R_{AA}$ decreases significantly  even for $L$=4, 5, 6 fm, indicating
that leading hadrons can originate from partons with significant
in-medium pathlength. 
Fig.~\ref{fig8} is consistent with the
statement in~\cite{Salgado:2003gb} that the suppression of hadrons
at RHIC can be accounted for by $\hat q=1$ ${\rm GeV}^2/{\rm fm}$
and $L\sim$ 6 fm. In the present study, we find a significantly larger
transport coefficient since the relevant in-medium pathlength
turns out to be much smaller. Consistent with the statement made above, 
if the transport coefficient is increased to 
$\hat q=10$ ${\rm GeV}^2/{\rm fm}$, partons with
in-medium pathlength more than
3 fm make a negligible contribution to
the nuclear modification factor, see Fig.~\ref{fig8}.
This is consistent with the model discussed in Ref.~\cite{Muller:2002fa}.
Effectively, with $\hat q=10$ ${\rm GeV}^2/{\rm fm}$
the results from RHIC can be reproduced
with a fixed path length $L\sim 2$ fm, which is 
clearly smaller than $\langle L_{\rm geom}\rangle =5.2$ fm (see also
\cite{Dainese:2003wq}). For $L\sim 2$ fm, 
the average momentum transfer acquired by the hard partons is of the
order $\sqrt{\hat qL}\sim 4.5$ GeV. Thus, the direction of partons
with $q_{T_i}\lsim 5$ GeV is randomized; they
become part of the medium.
This suggests that hadrons with $p_T\lsim 2-3$ GeV, which form the
bulk of the multiplicity, originate from decoupling of the strongly
interacting system. This observation is 
consistent with the results in \cite{Eskola:2002tu}.

%

\begin{figure}[h]
\vspace{-0.5cm}
\centerline{
\epsfysize=7cm\epsffile{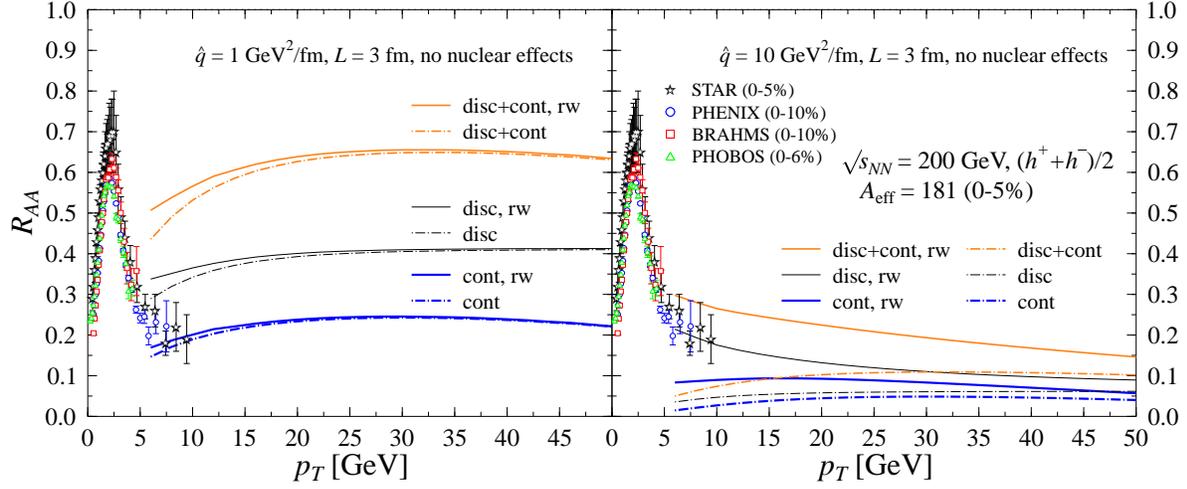} }
\caption[a]{\protect \small (Colour online)
The relative contributions of the discrete
and continuous parts of the quenching weight (\ref{eq6}) to the
nuclear modification factor $R_{AA}$ at RHIC for fixed 
in-medium path length $L=3$ fm and for $\hat q=1$  ${\rm GeV}^2/{\rm fm}$ 
(left panel) and $\hat q=10$  ${\rm GeV}^2/{\rm fm}$ (right panel),
respectively. The data are the same as in Fig.~\ref{fig3}.}
\label{fig9}
\end{figure}

We finally check to what extent the effects of parton energy
loss on $R_{AA}$ are due to the discrete or the continuous part
of the quenching weight (\ref{eq6}).  The discrete part $p_0$
depends for fixed  $L$ and $\hat q$ only on the parton 
species, see Fig.~\ref{fig2}.  The non-reweighted contribution of $p_0$
to $R_{AA}$ therefore interpolates between the values $p_0^{\rm gluon}$
and $p_0^{\rm quark}$ as a function of $p_T$. 
In Fig.~\ref{fig9} we fix $L=3$ fm,
and show separately the discrete and continuous contribution
to the nuclear modification factor $R_{AA}$ at RHIC for
$\hat q=1$ and 10 ${\rm GeV}^2/{\rm fm}$. No other nuclear effects
are included in Fig.~\ref{fig9}. 
The high-$p_T$ hadron spectrum at RHIC is dominated by quark production, 
and thus the non-reweighted discrete component of $R_{AA}$ is very close to 
$p_0^{\rm quark}$.
%

\begin{figure}[h]
\vspace{-0.5cm}
\centerline{
\epsfysize=7cm\epsffile{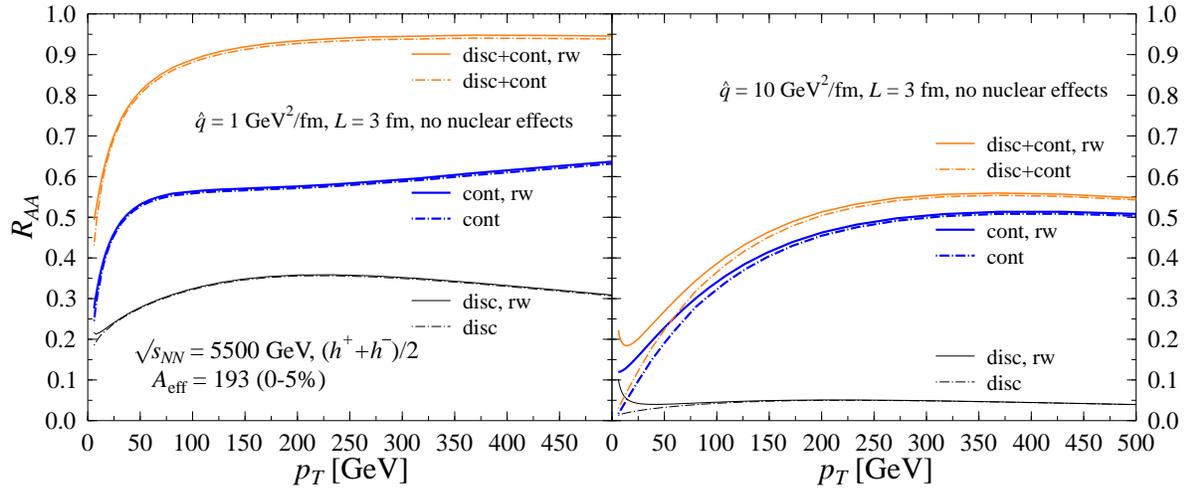} }
\caption[a]{\protect \small Same as in Fig.~\ref{fig9} for the LHC.}
\label{fig10}
\end{figure}

At the LHC, for $\hat q=1$ and 10 ${\rm GeV}^2/{\rm fm}$,
the suppression is weaker than at RHIC, see Fig.~\ref{fig10} and 
Sec.~\ref{sec4}. The contribution of the discrete part to $R_{AA}$ is
smaller than at RHIC since a significant part of the hadron production
is now of gluonic origin. [The slight decrease of the discrete 
contribution at $p_T\gsim 200$ GeV reflects the relative shape
of the KKP quark and gluon fragmentation functions at very high $z$.]
The dominant contribution to $R_{AA}$
comes from the continuous part of the quenching weight. Similar to RHIC, 
the increasing slope of the partonic spectra at high $p_T$ tames the 
growth of $R_{AA}$.

\begin{thebibliography}{99}

%
\bibitem{Adcox:2001jp}
K.~Adcox {\it et al.}  [PHENIX Collaboration],
Phys.\ Rev.\ Lett.\  {\bf 88} (2002) 022301
[arXiv:nucl-ex/0109003].
%
\bibitem{Adler:2003au}
S.~S.~Adler {\it et al.}  [PHENIX Collaboration],
Phys.\ Rev.\ C {\bf 69} (2004) 034910
[arXiv:nucl-ex/0308006].
%
\bibitem{Adler:2002xw}
C.~Adler {\it et al.} [STAR Collaboration],
Phys.\ Rev.\ Lett.\  {\bf 89} (2002) 202301
[arXiv:nucl-ex/0206011].
%
\bibitem{Adams:2003kv}
J.~Adams {\it et al.}  [STAR Collaboration],
Phys.\ Rev.\ Lett.\  {\bf 91} (2003) 172302
[arXiv:nucl-ex/0305015].
%
\bibitem{Back:2003qr}
B.~B.~Back {\it et al.}  [PHOBOS Collaboration],
Phys.\ Lett.\ B {\bf 578} (2004) 297
[arXiv:nucl-ex/0302015].
%
\bibitem{Arsene:2003yk}
I.~Arsene {\it et al.}  [BRAHMS Collaboration],
Phys.\ Rev.\ Lett.\  {\bf 91} (2003) 072305
[arXiv:nucl-ex/0307003].
%
\bibitem{Adler:2002tq}
C.~Adler {\it et al.}  [STAR Collaboration],
Phys.\ Rev.\ Lett.\  {\bf 90} (2003) 082302
[arXiv:nucl-ex/0210033].
%
\bibitem{Adler:2003ii}
S.~S.~Adler {\it et al.}  [PHENIX Collaboration],
Phys.\ Rev.\ Lett.\  {\bf 91} (2003) 072303
[arXiv:nucl-ex/0306021].
%
\bibitem{Adams:2003im}
J.~Adams {\it et al.}  [STAR Collaboration],
Phys.\ Rev.\ Lett.\  {\bf 91} (2003) 072304
[arXiv:nucl-ex/0306024].
%
\bibitem{Wiedemann:2004wp}
U.~A.~Wiedemann,
arXiv:hep-ph/0402251.
%
\bibitem{Gyulassy:1993hr}
M.~Gyulassy and X.~N.~Wang,
Nucl.\ Phys.\ B {\bf 420} (1994) 583
[arXiv:nucl-th/9306003].
%
\bibitem{Baier:1996sk}
R.~Baier, Y.~L.~Dokshitzer, A.~H.~Mueller, S.~Peigne and D.~Schiff,
Nucl.\ Phys.\ B {\bf 484} (1997) 265
[arXiv:hep-ph/9608322].
%
\bibitem{Zakharov:1997uu}
B.~G.~Zakharov,
JETP Lett.\  {\bf 65} (1997) 615
[arXiv:hep-ph/9704255].
%
\bibitem{Wiedemann:2000za}
U.~A.~Wiedemann,
Nucl.\ Phys.\ B {\bf 588} (2000) 303
[arXiv:hep-ph/0005129].
%
\bibitem{Gyulassy:2000er}
M.~Gyulassy, P.~Levai and I.~Vitev,
Nucl.\ Phys.\ B {\bf 594} (2001) 371
[arXiv:nucl-th/0006010].
%
\bibitem{Wang:2001if}
X.~N.~Wang and X.~f.~Guo,
Nucl.\ Phys.\ A {\bf 696} (2001) 788
[arXiv:hep-ph/0102230].
%
\bibitem{Wang:2003aw}
X.~N.~Wang,
Phys.\ Lett.\ B {\bf 579}, 299 (2004)
[arXiv:nucl-th/0307036].
%
\bibitem{Gyulassy:2004vg}
M.~Gyulassy,
arXiv:nucl-th/0403032.
%
\bibitem{Eskola:2002kv}
K.~J.~Eskola and H.~Honkanen,
Nucl.\ Phys.\ A {\bf 713} (2003) 167
[arXiv:hep-ph/0205048].
%
\bibitem{Eskola:1998df}
K.~J.~Eskola, V.~J.~Kolhinen and C.~A.~Salgado,
Eur.\ Phys.\ J.\ C {\bf 9} (1999) 61
[arXiv:hep-ph/9807297].
%
\bibitem{Salgado:2003gb}
C.~A.~Salgado and U.~A.~Wiedemann,
Phys.\ Rev.\ D {\bf 68} (2003) 014008
[arXiv:hep-ph/0302184].
%
\bibitem{Baier:2002tc}
R.~Baier,
Nucl.\ Phys.\ A {\bf 715} (2003) 209
[arXiv:hep-ph/0209038].
%
\bibitem{Baier:2001yt}
R.~Baier, Y.~L.~Dokshitzer, A.~H.~Mueller and D.~Schiff,
JHEP {\bf 0109} (2001) 033
[arXiv:hep-ph/0106347].
%
\bibitem{Salgado:2002cd}
C.~A.~Salgado and U.~A.~Wiedemann,
Phys.\ Rev.\ Lett.\  {\bf 89} (2002) 092303
[arXiv:hep-ph/0204221].
%
\bibitem{Baier:1998yf}
R.~Baier, Y.~L.~Dokshitzer, A.~H.~Mueller and D.~Schiff,
Phys.\ Rev.\ C {\bf 58} (1998) 1706
[arXiv:hep-ph/9803473].
%
\bibitem{Wiedemann:2000tf}
U.~A.~Wiedemann,
Nucl.\ Phys.\ A {\bf 690} (2001) 731
[arXiv:hep-ph/0008241].
%
\bibitem{Gyulassy:2001nm}
M.~Gyulassy, P.~Levai and I.~Vitev,
Phys.\ Lett.\ B {\bf 538} (2002) 282
[arXiv:nucl-th/0112071].
%
\bibitem{Salgado:2003rv}
C.~A.~Salgado and U.~A.~Wiedemann,
arXiv:hep-ph/0310079.
%
\bibitem{Dainese:2003wq}
A.~Dainese  [ALICE Collaboration],
Eur.\ Phys.\ J.\ C {\bf 33} (2004) 495
[arXiv:nucl-ex/0312005];
A.~Dainese,
arXiv:nucl-ex/0311004.
%
\bibitem{Eskola:2001bf}
K.~J.~Eskola, P.~V.~Ruuskanen, S.~S.~R\"as\"anen and K.~Tuominen,
Nucl.\ Phys.\ A {\bf 696} (2001) 715
[arXiv:hep-ph/0104010].
%
\bibitem{Lai:1999wy}
H.~L.~Lai {\it et al.}  [CTEQ Collaboration],
Eur.\ Phys.\ J.\ C {\bf 12} (2000) 375
[arXiv:hep-ph/9903282].
%
\bibitem{Kniehl:2000fe}
B.~A.~Kniehl, G.~Kramer and B.~Potter,
Nucl.\ Phys.\ B {\bf 582} (2000) 514
[arXiv:hep-ph/0010289].
%
\bibitem{Wang:1996yh}
X.~N.~Wang, Z.~Huang and I.~Sarcevic,
Phys.\ Rev.\ Lett.\  {\bf 77} (1996) 231
[arXiv:hep-ph/9605213].
%
\bibitem{Armesto:2000xh}
N.~Armesto and C.~Pajares,
Int.\ J.\ Mod.\ Phys.\ A {\bf 15} (2000) 2019
[arXiv:hep-ph/0002163].
%
\bibitem{EHNRRT}
K.~J.~Eskola, H.~Honkanen, H.~Niemi, V.~P.~Ruuskanen, S.~S.~R\"as\"anen
and K.~Tuominen, 
in preparation.
%
\bibitem{Eskola:1999fc}
K.~J.~Eskola, K.~Kajantie, P.~V.~Ruuskanen and K.~Tuominen,
Nucl.\ Phys.\ B {\bf 570} (2000) 379
[arXiv:hep-ph/9909456].
%
\bibitem{Vitev:2004gn}
I.~Vitev,
arXiv:nucl-th/0404052.
%
\bibitem{Adil:2004cn}
A.~Adil and M.~Gyulassy,
arXiv:nucl-th/0405036.
%
\bibitem{Wang:2004yv}
X.~N.~Wang,
arXiv:nucl-th/0405029.
%
\bibitem{Adler:2003kg}
S.~S.~Adler {\it et al.}  [PHENIX Collaboration],
Phys.\ Rev.\ Lett.\  {\bf 91} (2003) 172301.
%
\bibitem{Adams:2003am}
J.~Adams {\it et al.}  [STAR Collaboration],
arXiv:nucl-ex/0306007.
%
\bibitem{dA} S.~S.~Adler {\it et al.}  [PHENIX Coll.], arXiv:nucl-ex/0306021;
J.~Adams  [STAR Coll.], arXiv:nucl-ex/0306024;
B.~B.~Back {\it et al.} [PHOBOS Coll.], arXiv:nucl-ex/0306025;
I.~Arsene {\it et al.} [BRAHMS Coll.],
arXiv:nucl-ex/0307003.
%
\bibitem{Back:2004ra}
B.~B.~Back  [PHOBOS Collaboration],
arXiv:nucl-ex/0405003.
%
\bibitem{pAprelim}
Talks by Takao Sakaguchi [PHENIX Collaboration] and
Marco van Leeuwen [STAR Collaboration], 2004 AGS \& RHIC Annual
Users' Meeting, May 10-14, 2004, 
http://www.phenix.bnl.gov/~enterria/agsrhic04\_hipt/
%
\bibitem{Molnar:2001ux}
D.~Molnar and M.~Gyulassy,
Nucl.\ Phys.\ A {\bf 697} (2002) 495
[Erratum-ibid.\ A {\bf 703} (2002) 893]
[arXiv:nucl-th/0104073].
%
\bibitem{Dainese:2004te}
A.~Dainese, C.~Loizides and G.~Paic,
arXiv:hep-ph/0406201.
%
\bibitem{Drees:2003zh}
A.~Drees, H.~Feng and J.~Jia,
arXiv:nucl-th/0310044.
%
\bibitem{Muller:2002fa}
B.~Muller,
Phys.\ Rev.\ C {\bf 67} (2003) 061901
[arXiv:nucl-th/0208038].
%
\bibitem{Eskola:2002tu}
K.~J.~Eskola, H.~Niemi, P.~V.~Ruuskanen and S.~S.~R\"as\"anen,
Nucl.\ Phys.\ A {\bf 715} (2003) 561
[arXiv:nucl-th/0210005].

\end{thebibliography}
\end{document}